\documentclass{article}
\usepackage{spconf,amsmath,graphicx}
\usepackage{amssymb}
\usepackage{booktabs}
\usepackage{multirow}

\title{Multi-dimensional frequency dynamic convolution with confident mean teacher for sound event detection }
\name{Shengchang Xiao$^{1,2}$ , Xueshuai Zhang$^{1,\star}$ \thanks{$^\star$Corresponding author}, Pengyuan Zhang$^{1,2}$}
\address{
  $^1$Key Laboratory of Speech Acoustics and Content Understanding, Institute of Acoustics, 
  \\Chinese Academy of Sciences\\
  $^2$University of Chinese Academy of Sciences}

\begin{document}
%
\maketitle
\begin{abstract}
Recently, convolutional neural networks (CNNs) have been widely used in sound event detection (SED). However, traditional convolution is deficient in learning time-frequency domain representation of different sound events. To address this issue, we propose multi-dimensional frequency dynamic convolution (MFDConv), a new design that endows convolutional kernels with frequency-adaptive dynamic properties along multiple dimensions. MFDConv utilizes a novel multi-dimensional attention mechanism with a parallel strategy to learn complementary frequency-adaptive attentions, which substantially strengthen the feature extraction ability of convolutional kernels. Moreover, in order to promote the performance of mean teacher, we propose the confident mean teacher to increase the accuracy of pseudo-labels from the teacher and train the student with high confidence labels. Experimental results show that the proposed methods achieve 0.470 and 0.692 of PSDS1 and PSDS2 on the DESED real validation dataset.
\end{abstract}
\section{Introduction}
Sound event detection (SED) task aims at detecting specific sound events present in audio clips and it has been  widely used in medical, wearable devices and intelligent security area. Recently, semi-supervised SED has attracted increasing research interest in the Detection and Classification of Acoustic Scenes and Events (DCASE) challenge Task4 \cite{turpault2019sound}. 

With the development of deep learning (DL), SED has adopted various DL methods and achieved great success. Among these methods, CNN \cite{salamon2017deep} is commonly used to extract the high dimensional representation from audio features. However, the feature extraction ability of basic convolution is limited. To handle this limitation, there have been several attempts to incorporate attention mechanism into convolutional blocks including SENet \cite{hu2018squeeze}, SKNet \cite{li2019selective} and CBAM \cite{woo2018cbam}. Recently, dynamic convolution \cite{yang2019condconv,chen2020dynamic} which aggregates multiple parallel convolutional kernels dynamically based on their attentions has become popular in optimizing efficient CNNs. Despite its 
performance improvement, dynamic convolution has a crucial limitation that only one dimension (convolutional kernel number) is endowed with dynamic property while the other dimensions are overlooked \cite{li2022omni}.  

In addition, these methods are primarily designed for image data and not exactly compatible with time-frequency spectrogram. Specifically, image is translation invariant on both dimensions while the time-frequency spectrogram is not translation invariant on frequency dimension. To address this problem, frequency dynamic convolution (FDConv) \cite{nam2022frequency} is proposed to release translation equivariance of convolution on frequency dimension. FDConv applies frequency-adaptive kernels to strengthen frequency-dependency on convolution and achieves competitive results on SED task.

Another challenge for SED task is the lack of well annotated datasets. To solve this problem, various semi-supervised learning (SSL) \cite{laine2016temporal,sajjadi2016regularization,miyato2018virtual} methods are proposed to exploit unlabelled data. Among these methods, mean teacher (MT) \cite{tarvainen2017mean} has achieved promising SED performance.  In order to further promote the performance, some improved MT methods are proposed. Guided learning designed a teacher for audio tagging (AT) to guide a student for SED \cite{lin2020guided}. Task-aware mean teacher utilize a CRNN with multi-branch structure to solve the SED and AT tasks differently \cite{yan2020task}. Interpolation consistency training (ICT) \cite{verma2019interpolation} and shift consistency training (SCT) \cite{koh2021sound} are proposed to exploit large amount of unlabeled in-domain data efficiently. However, these methods can't solve the problem that the inaccurate pseudo-label obtained from the teacher will lead to confirmation bias and wrong training directions \cite{liu2022perturbed}.

In this paper, we propose multi-dimensional frequency dynamic convolution (MFDConv) and confident mean teacher (CMT) to address the two challenges respectively. Firstly, in order to strengthen the feature extraction ability of FDConv, we extend the frequency-adaptive dynamic properties of convolutional kernels to more dimensions of the kernel space. MFDConv utilizes a novel multi-dimensional attention mechanism with a parallel strategy to learn these  frequency-adaptive attentions for convolutional kernels. We demonstrate that these attentions along different dimensions are complementary to each other and progressively applying them to the corresponding convolutional kernels can substantially improve the representation power of basic convolution. Secondly, to further promote the performance of MT, we introduce the confident mean teacher to solve the pseudo-label accuracy problem. In particular, we perform the weak-strong thresholding and event-specific median filter on the teacher prediction to improve the precision of pseudo-label. Furthermore, we adopt the confidence-weighted BCE loss instead of MSE loss for consistency training to help the student model train with high confidence pseudo-label. Experimental results on the DCASE2021 Task4 dataset validate the superior performance of proposed methods.

\section{Method}
\subsection{Dynamic Convolution} 

A basic convolution can be denoted as $\boldsymbol{y}=\boldsymbol{W}*\boldsymbol{x}+\boldsymbol{b}$, where $\boldsymbol{W}$ and $\boldsymbol{b}$ are weight and bias of a basis kernel. For dynamic convolution \cite{chen2020dynamic}, it aggregates multiple parallel convolution kernels dynamically based on their attentions which are input-dependent. Mathematically, the dynamic convolution can be defined as:
\begin{equation}
\begin{aligned}
&y=(\sum\limits_{i=1}^n\alpha_{wi}W_{i})*x\\
&\alpha_{w i}=\pi_{w i}(x)\\
\end{aligned}
\end{equation}
where $x\in\mathbb{R}^{T\times F\times c_{in}}$ and $y\in\mathbb{R}^{T\times F\times c_{out}}$ denote the input features and the output features; $W_{i}\in\mathbb{R}^{k\times k\times c_{in}\times c_{out} }$ denotes the $i^{th}$ convolutional kernel; $\alpha_{wi}\in\mathbb{R}$ is the attention weight for the the $i^{th}$ convolutional kernel, which is computed by the attention function $\pi_{wi}(x)$ conditioned on the input features. For simplicity, the bias term is omitted.

\subsection{Multi-dimensional Frequency Dynamic Convolution} 

In fact, for $n$ convolutional kernels, the corresponding kernel space has 4 dimensions including the kernel number $n$, the input channels $c_{in}$, the output channels $c_{out}$ and the spatial kernel size $k\times k$. However, dynamic convolution only endow convolutional kernels with the dynamic property along one dimension (the convolutional kernel number) of the kernel space, while the other three dimensions are ignored. 
The attention function $\pi_{wi}(x)$ calculate one attention weight for the convolutional kernel $W_{i}$, which means that all its filters have the same attention value for the input. 

Therefore, we extend the frequency-adaptive dynamic properties of convolutional kernels to more dimensions of the kernel space. Our multi-dimensional frequency dynamic convolution (MFDConv) can be defined as follows:
\begin{gather}
y=(\sum\limits_{i=1}^n\alpha_{wi}(f)\odot\alpha_{fi}(f)\odot\alpha_{ci}(f)\odot W_{i})*x \nonumber\\
\alpha_{w i}(f)=\pi_{w i}(x,f) \\
\alpha_{f i}(f)=\pi_{f i}(x,f) \nonumber\\
\alpha_{c i}(f)=\pi_{c i}(x,f)\nonumber
\end{gather}
\begin{figure}[t]
  \centering
  \includegraphics[width=\linewidth]{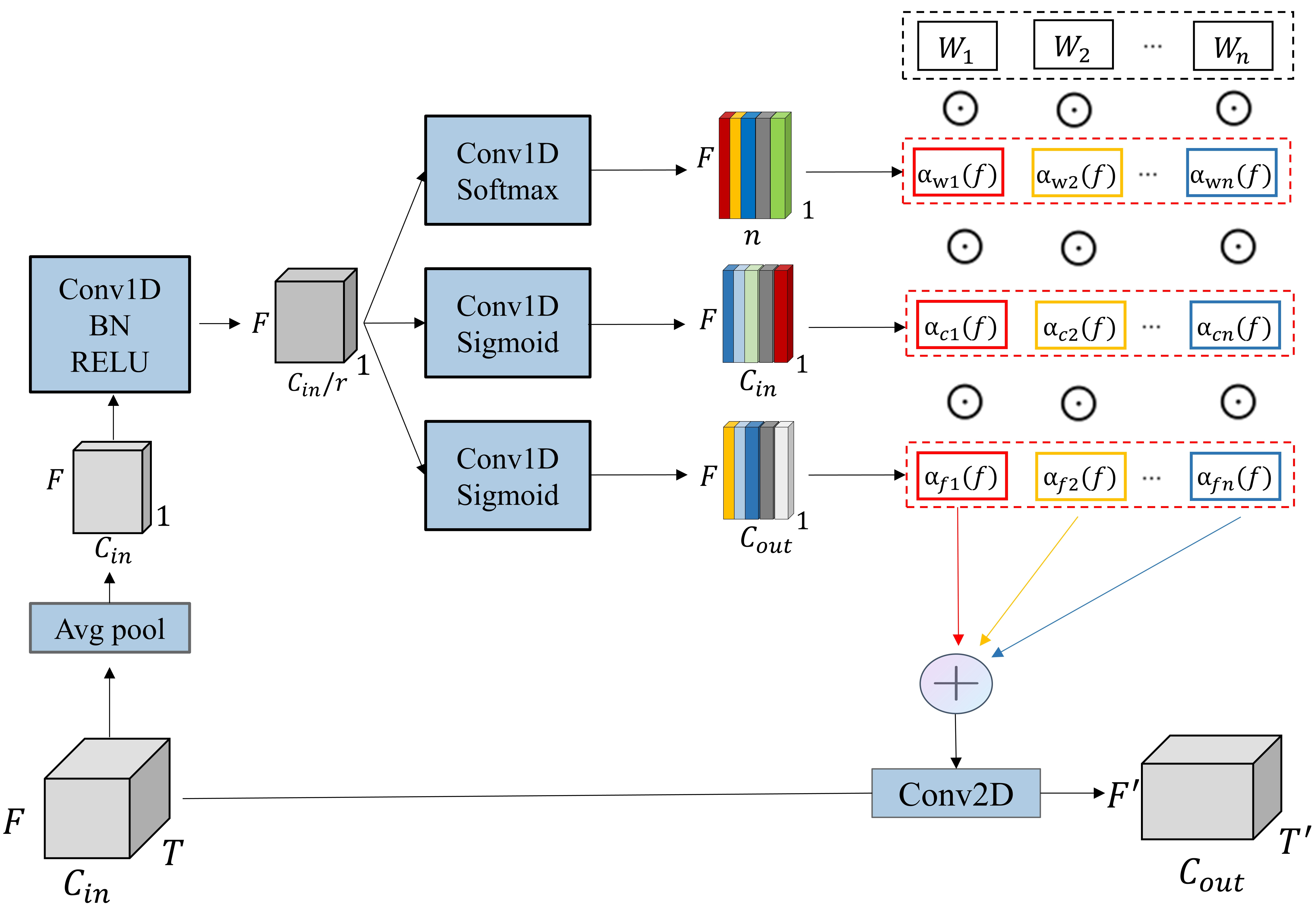}
  \caption{\textit{The illustration of multi-dimensional frequency dynamic convolution operation. $F$ and $T$ denotes the frequency dimension and time dimension; $C_{in}$ and $C_{out}$ denote the input channel and output channel. $r$ is the reduction ratio and $n$ is the number of basic kernels. $\alpha_{w i}(f)$, $\alpha_{ci}(f)$ and $\alpha_{fi}(f)$ denote the frequency-adaptive attention weights.}}
  \label{fig:1}
\end{figure}
where $\alpha_{w i}(f)$ is the frequency-adaptive attention weights for the convolutional kernel $W_{i}$;  $\alpha_{fi}(f)\in\mathbb{R}^{c_{out}}$, $\alpha_{ci}(f)\in\mathbb{R}^{c_{in}}$ denote newly introduced frequency-adaptive attentions computed along the output channel dimension and the input channel dimension; $\odot$ denotes the multiplication operations along different dimensions of the kernel space. $\alpha_{w i}(f)$, $\alpha_{fi}(f)$ and $\alpha_{ci}(f)$ are computed by a multi-head attention module consisting of $\pi_{w i}(x,f)$, $\pi_{f i}(x,f)$ and $\pi_{c i}(x,f)$. Note that the dimension spatial kernel size $k\times k$ is not used.

In MFDConv, for the the convolutional kernel $W_{i}$: (1) $\alpha_{ci}(f)$ assigns frequency-adaptive attention weights to $c_{in}$ channels; (2) $\alpha_{fi}(f)$ assigns frequency-adaptive attention weights to $c_{out}$ channels; (3) $\alpha_{wi}(f)$ assigns frequency-adaptive attention weights to the whole convolutional kernel. In theory, the three frequency-adaptive attention are complement to each other and applying them to the corresponding convolutional kernels can substantially strengthen the representation capability of basic convolution. The architecture of MFDConv is shown in Fig~\ref{fig:1}.

For the implementation, we adopt the SE module \cite{hu2018squeeze} to extract frequency-adaptive attention weights. The difference is that MFDConv have multiple heads to compute $\pi_{w i}(x,f)$, $\pi_{f i}(x,f)$ and $\pi_{c i}(x,f)$ respectively. In particular, we first apply average pool along time dimension to squeeze the input into a feature map with the shape of $F\times{c_{in}}$. Subsequently, a 1D convolution block squeeze the feature map to a lower dimensional space with reduction ratio $r$. Then there are 3 branches of 1D convolution layer with the output size of $F\times n$, $F\times{c_{out}}$ and $F\times c_{in}$. Finally, a softmax or sigmoid function is applied to obtain the normalized frequency-adaptive attention weights $\alpha_{wi}(f)$, $\alpha_{fi}(f)$ and $\alpha_{ci}(f)$.

\subsection{Confident Mean Teacher} 

\begin{figure}[t]
  \centering
  \includegraphics[width=\linewidth]{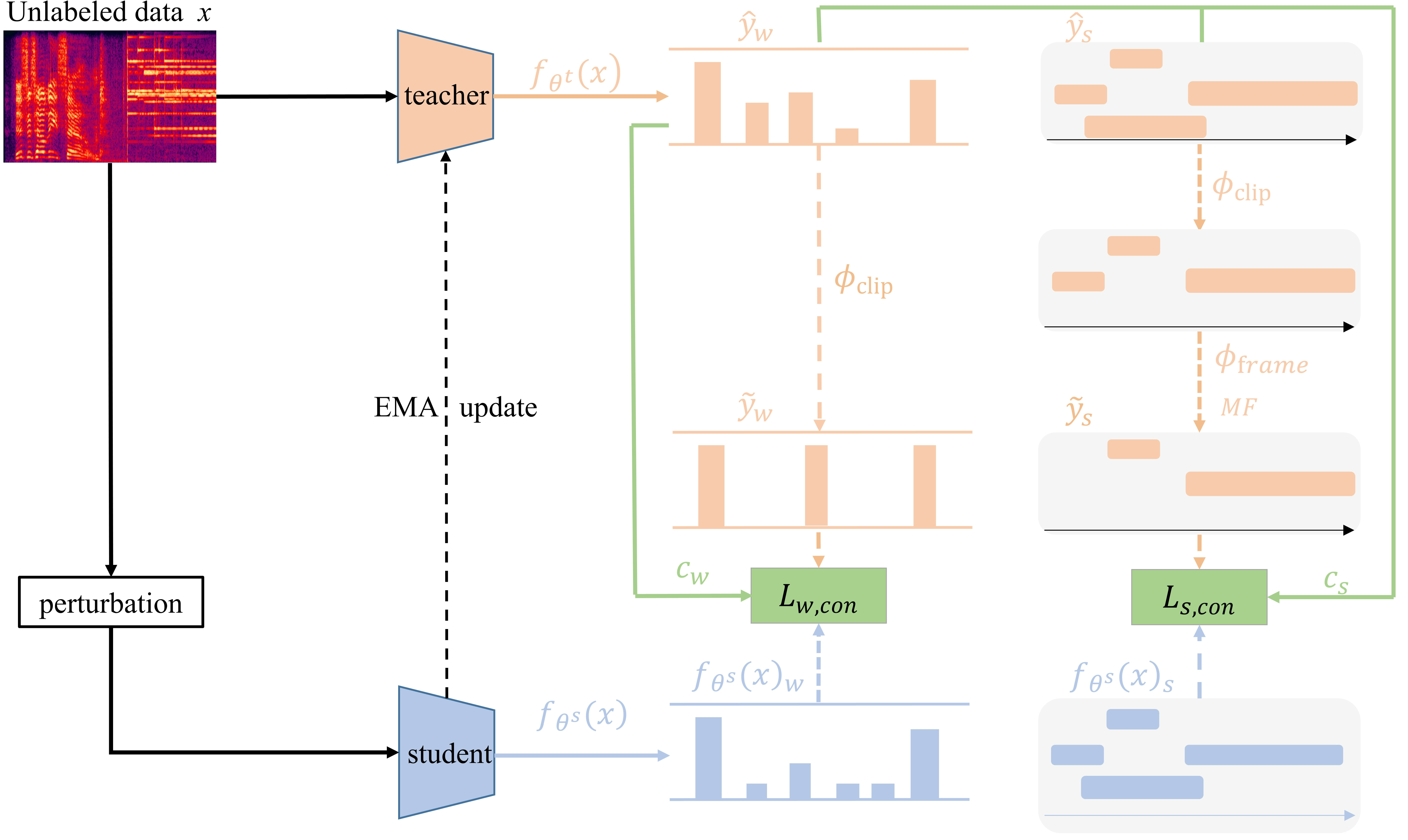}
  \caption{\textit{The structure of confident mean teacher. $\tilde{y}_{w}$ and $\tilde{y}_{s}$ denote the clip-wise and frame-wise prediction of teacher. $f_{\theta^{s}}(x)_{w}$ and $f_{\theta^{s}}(x)_{s}$ denote the prediction of student. $\hat{y}_{w}$ and $\hat{y}_{s}$ denote the corrected pseudo-labels.$L_{w,con}$ and $L_{s,con}$ denote the clip-wise and frame-wise consistency loss. $c_{w}$ and $c_{s}$ denote the confidence weight .} }
  \label{fig:2}
\end{figure}

Traditional mean teacher can be seriously affected by the inaccurate predictions of unlabeled data. 
Therefore, we propose the confident mean teacher (CMT) method to address the pseudo-label accuracy problem. The core idea of CMT is to correct inaccurate predictions from the teacher by post-processing operations and train the student with high confidence labels. The structure of CMT is shown in Fig~\ref{fig:2}.

In particular, we first obtain the clip-wise prediction $\hat{y}_{w}\in[0,1]^{K}$ and frame-wise prediction $\hat{y}_{s}\in[0,1]^{T\times K}$ from the teacher model. $T$ and $K$ denote the frame number and sound event class number. Then we set a clip-wise threshold $\phi_{clip}$. If $\hat{y}_{w}>\phi_{clip}$, $\hat{y}_{w}$ is assigned to 1. Otherwise, $\hat{y}_{w}$ is assigned to 0. If $\hat{y}_{s}<\phi_{clip}$, $\hat{y}_{s}$ is assigned to 0. In addition to weak threshold, we also set the frame-wise threshold $\phi_{frame}$. If $\hat{y}_{s}>\phi_{frame}$, $\hat{y}_{s}$ is assigned to 1. Otherwise, $\hat{y}_{s}$ is assigned to 0. After strong threshold, we smooth the frame-wise prediction $\hat{y}_{s}$ with event-specific median filters. These steps can be denoted as follows:
\begin{gather}
\tilde{y}_{w}(k)=\mathbb{I}(\hat{y}_{w}(k)>\phi_\textrm{clip})\\
\tilde{y}_{s}(t,k)=MF(\mathbb{I}(\hat{y}_{w}(k)>\phi_\textrm{clip})\mathbb{I}(\hat{y}_{s}(t,k)>\phi_\textrm{frame}))
\end{gather}
where $\tilde{y}_{w}$ and $\tilde{y}_{s}$ denote the clip-wise pseudo-label and frame-wise pseudo-label respectively; $\mathbb{I}(.)$ is the indicator function. $MF$ denotes the median filters. Compared to the initial prediction, the pseudo-label is more reliable and the student model is more difficult to overfit the pseudo-label. Furthermore, we apply confidence weight to the consistency loss according to the prediction probabilities. The consistency loss consists of clip-wise consistency loss $\ell_{w,con}$ and frame-wise consistency loss $\ell_{s,con}$. They can be defined as follows:
\begin{equation}
    \ell_{w,con}=\frac{1}{|K|}\sum_{k\in K}c_{w}(k)\ell(\tilde{y}_{w}(k),f_{\theta^{s}}(x)_{w}(k))
\end{equation}
\begin{equation}
    \ell_{s,con}=\frac{1}{|\Omega|}\sum_{t,k\in\Omega}c_{s}(t,k)\ell(\tilde{y}_{s}(t,k),f_{\theta^{s}}(x)_{s}(t,k))
\end{equation}
where $K$ is the sound event class number and $\Omega$ is the frame-wise probability map of size $T\times K$. $\ell(.)$ denotes the BCE loss between pseudo-label and student prediction. $\tilde{y}_{w}(k)$ and $\tilde{{y}_{s}}(t,k)$ denote the clip-wise pseudo-label at class $k$ and frame-wise pseudo-label at the specific frame and class $(t,k)$; $f_{\theta^{s}}(x)_{w}(k)$ and $f_{\theta^{s}}(x)_{s}(t,k)$ denote the clip-wise prediction at $k$ and frame-wise prediction at $(t,k)$ from the student model $\theta^{s}$; $c_{w}(k)$ and $c_{s}(t,k)$ denote the clip-wise prediction confidence at $k$ and the frame-wise prediction confidence at $(t,k)$ from the teacher model $\theta^{t}$. The confidence weight is computed as:
\begin{equation}
    c_{w}(k)=\hat{y}_{w}(k)\mathbb{I}(\tilde{y}_{w}(k)=1)
\end{equation}
\begin{equation}
    c_{s}(t,k)=\hat{y}_{s}(t,k)\hat{y}_{w}(k)\mathbb{I}(\tilde{y}_{s}(t,k)=1)
\end{equation}
The weighted consistency loss can train the student model with high confidence pseudo-labels and reduce the impact of inaccurate pseudo-label.

\section{Experiments and results}
\subsection{Dataset}
Our experiments are conducted on the dataset of Task4 in the DCASE2021. The development set contains three types of training data: weakly labeled data (1578 clips), synthetic strongly labeled data (10000 clips) and unlabeled in domain data (14412 clips). The validation set (1168 clips) is used for evaluation.
We extract the Log-Mel spectrogram on 16kHz audio with 128 mel frequency bins as features.

\subsection{Experimental Setups}
Our baseline model is the CRNN architecture \cite{cakir2017convolutional}. And we set the same hyper-parameters as \cite{nam2022frequency} for comparison purposes. For our MFDConv, the reduction ratio $r$ is set to 4 and the basic kernel number $n$ is set to 4. For CMT, the clip-wise threshold $\phi_{{clip}}$ and the frame-wise threshold $\phi_{{frame}}$ are both set to 0.5.
The poly-phonic sound event detection scores (PSDS) \cite{bilen2020framework} and Collar-based F1 is used to evaluate the performance of SED models. 


\subsection{Comparison of Different Convolution}
We first compare the performance of baseline with different convolution methods including dynamic convolution (Dyconv) \cite{chen2020dynamic}, frequency dynamic convolution (FDConv) \cite{nam2022frequency}, omni-dimensional dynamic convolution (ODConv) \cite{li2022omni} and our MFDConv. The experiments in this section adopt basic mean teacher method for semi-supervised learning.

\begin{table}[h]
\caption{\textit{SED performance comparison between models using different dynamic convolution on the validation set}}
\label{tab:1}
\centering
\begin{tabular}{lccc}
\toprule
\multicolumn{1}{l}{\textbf{Model}}    & \multicolumn{1}{l}{\textbf{PSDS1}} & \multicolumn{1}{l}{\textbf{PSDS2}} & \multicolumn{1}{l}{\textbf{F1-score}}  \\
\midrule
\multicolumn{1}{l}{Baseline} & 0.418                     & 0.640                     & 0.519                                \\
\midrule
+DyConv    & 0.439                     & 0.660                     & 0.525                                \\
+FDConv   & 0.450                     & 0.667                     & 0.533                                \\
+ODConv   & 0.445                    & 0.664                     & 0.528                                \\
+MFDConv  & \textbf{0.461}                     & \textbf{0.680}                     & \textbf{0.542}                  \\
\bottomrule
\end{tabular}
\end{table}
The results are shown in Table~\ref{tab:1}. Compared with the baseline, each dynamic convolution method can improve the SED performance. This prove that aggregating multiple parallel convolution kernels dynamically can promote the ability of feature extraction. In addition, we can observe that the frequency dynamic convolution outperforms the basic dynamic convolution. This is because FDConv applies frequency-adaptive attention weights to each kernel and this dynamic property is more consistent with SED task. Moreover, ODConv performs better than DyConv because ODConv learns extra dynamic attention along multiple dimensions. Combining the multi-dimensional attention with frequency-adaptive 
property, our MFDConv methods achieves the best results. This demonstrates that the proposed MFDConv can further strengthen the representation capability of basic convolution for SED task.

\subsection{Dependency of Different Dimensional Attention}
\begin{table}[t]
\caption{\textit{Dependency of different dimensional attention.} }
\label{tab:2}
\centering
\resizebox{\linewidth}{!}{
\begin{tabular}{lcccccc}
\toprule
\multicolumn{1}{l}{\textbf{Model}}  & \multicolumn{1}{l}{\textbf{$\alpha_{c i}$}}  & \multicolumn{1}{l}{\textbf{$\alpha_{f i}$}} & \multicolumn{1}{l}{\textbf{$\alpha_{w i}$}} & \multicolumn{1}{l}{\textbf{PSDS1}} & \multicolumn{1}{l}{\textbf{PSDS2}} & \multicolumn{1}{l}{\textbf{F1-score}}  \\
\midrule
Baseline &-&-&-& 0.418                     & 0.640                     & 0.519                                \\
\midrule
\multirow{7}{*}{+MFDConv}   &\checkmark&-&- & 0.427                     & 0.650                     & 0.521                                \\
  &-&\checkmark&- & 0.431                     & 0.654                     & 0.524                                \\
  &-&-&\checkmark & 0.439                    & 0.660                     & 0.525                               \\
  &\checkmark&\checkmark&- & 0.442                     & 0.662                     & 0.528                                \\
 &-&\checkmark&\checkmark & 0.446                    & 0.667                    &0.531                 \\
  &\checkmark&-&\checkmark & 0.443                     & 0.665                   & 0.530               \\
   &\checkmark&\checkmark&\checkmark & \textbf{0.461}                     & \textbf{0.680}                     & \textbf{0.542}                  \\
\bottomrule
\end{tabular}}
\end{table}
Note that MFDConv has three types of convolutional kernel attentions $\alpha_{w i}$, $\alpha_{c i}$ and $\alpha_{f i}$ computed along three dimensions of the kernel space. In order to investigate the complementarity of these attentions, we perform a set of ablative experiments with different combinations of them. The results are shown in Table~\ref{tab:2}.
It can be observed that the combinations of any two attentions outperform any single attention. And our MFDConv with all three dimensional attentions achieves the best results. The results indicate that the three types of convolutional kernel attention are complementary to each other.

\subsection{Performance of Confident Mean Teacher}
In this section, we compare the performance of our confident mean teacher (CMT) method with basic mean teacher (MT).  In particular, we use the baseline, FDConv and MFDConv architecture to evaluate CMT respectively. 
For consistency training, CMT adopts confidence weighted BCE loss instead of MSE loss in MT.
The results are shown in Table~\ref{tab:3}. For each model architecture, our CMT method performs better than MT. Compared with MT, CMT can modify the inaccurate predictions from the teacher model, which can help student model learn more accurately. Furthermore, the confidence weighted loss tends to focus on the prediction with high confidence and ignore the prediction with low confidence.
\begin{table}[h]
\caption{\textit{Comparison of CMT and MT} }
\label{tab:3}
\centering
\begin{tabular}{lcccc}
\toprule
\textbf{Model} &\textbf{SSL} & \textbf{PSDS1} & \textbf{PSDS2} & \textbf{F1-score}\\
\midrule
\multirow{2}{*}{baseline} & MT & 0.418                     & 0.640                     & 0.519                                \\
         &CMT      &   0.423    & {0.645}                     & {0.522}                               \\
\midrule
\multirow{2}{*}{+FDConv} &MT  & 0.450                     & 0.667                     & 0.533                                \\
              &CMT    & {0.457} & {0.675}                     & {0.536}                                \\
\midrule
\multirow{2}{*}{+MFDConv}&MT   & 0.461                     & 0.680                     & 0.542                                \\
        &CMT   &  {0.470}        & {0.692}                     & {0.548}                  \\
\bottomrule
\end{tabular}
\end{table}

\section{Conclusion}
In this paper, we propose multi-dimensional frequency dynamic convolution (MFDConv), a more generalized dynamic convolution design that endows convolutional kernels with frequency-adaptive dynamic properties along multi dimensions. 
In addition, to solve pseudo-label accuracy problem, we present the confident mean teacher (CMT) method to correct the erroneous predictions from the teacher model and replace the MSE loss with confidence-weighted BCE loss for consistency learning. By combining MFDConv and CMT, our approach achieves 0.470 of PSDS1 and 0.692 of PSDS2. In the future, we aim to explore other convolution and semi-supervised methods to further solve existing problems.

\section{Acknowledgements}
This work is funded by China Postdoctoral Science Foundation (NO.2022M72332)

\vfill\pagebreak

\bibliographystyle{IEEEbib}
\bibliography{strings,refs}

\end{document}